\documentclass[twocolumn]{IEEEtran}

\usepackage{amsmath, graphicx, algorithm, algorithmic, enumerate, color}
\usepackage[noadjust]{cite}
\usepackage[english]{babel}
\usepackage{flushend}

\begin{document}

\title{UAV-Aided Jamming for Secure Ground Communication with Unknown Eavesdropper Location}

\author{{Christantus O. Nnamani},
{Muhammad R. A. Khandaker}, \IEEEmembership{Senior Member, IEEE}, 
{and Mathini Sellathurai}, \IEEEmembership{Senior Member, IEEE}

\thanks{The authors are with the School of Engineering and Physical Sciences, Heriot-Watt University, Edinburgh EH14 4AS, UK.}
}
%\address[]{School of Engineering and Physical Sciences, Heriot-Watt University, Edinburgh EH14 4AS, United Kingdom (e-mail: \{con1, m.khandaker, m.sellathurai\}@hw.ac.uk)}

%\tfootnote{This work was supported in part by the EPSRC Project EP/P009670/1.}

\maketitle
%%%%%%%%%%%%%%%%%%%%%%%%%%%%%%%%%%%%%%%%%%%%%

\begin{abstract}

%Secured communication is a primordial requirement for technological and military exploits. It has become more stringent as the gains of technological growth advance the mode and means of information theft. Signal jamming, a physical layer protection strategy, is the most prominent keyless brute force method of limiting the information theft. Nevertheless, the obscurity of the eavesdropper makes it difficult to appropriately deliver the jamming signal practically. While several methods of delivering the jamming signal exit in literature, the unmanned aerial vehicle (UAV) technique seem to be more effective considering the obscurity of the theft agent and the  benefit of aerial vision. However, the evasiveness of the eavesdropper is still a primary concern to its deployment. 
This paper investigates unmanned aerial vehicle (UAV)-aided jamming technique for enabling physical layer keyless security in scenarios where the exact eavesdropper location is unknown. We assume that the unknown eavesdropper location is within an ellipse characterizing the coverage region of the transmitter. By  sequentially optimizing the transmit power, the flight path of the UAV and its jamming power, we aim at maximizing the average secrecy rate with arbitrary eavesdropper location. Simulation results demonstrate that the optimal flight path obtains better secrecy rate performance compared to that using direct UAV flight path encasing the transmitter and the legitimate receiver. Most importantly, even with the unknown eavesdropper location, we obtained a  secrecy  rate  that  is  comparable  to  a  scenario  when  the eavesdropper’s  location  is  known. However, the  average  secrecy rate with the unknown eavesdropper location varies depending on the proximity of the eavesdropper to the known location of the transmitter. We also observe that due to the UAV-aided jamming, the average secrecy rate stabilizes at some point even though the average received envelope power of the eavesdropper increases. This essentially demonstrates the effectiveness of the proposed scheme.
%We showed that the highest information content obtained by the eavesdropper is possible only when it is at the same position as the legitimate receiver, else the UAV jamming signal will continue to degrade its information even if it has better channel state.
\end{abstract}


\maketitle

%\begin{IEEEkeywords}
%Secure communication, Jamming, UAV, Optimization, Physical layer security.
%\end{IEEEkeywords}

\section{Introduction}
Protecting sensitive or confidential information is of paramount interest to most businesses/organizations – private, public, government, military or intelligence. In the event that such data/information is made public, these businesses/organizations may face legal or financial ramifications. At the very least, they will suffer loss of customer trust (e.g. companies, etc.); but in the worst case, it could lead to the complete annihilation of the organization (e.g. Military, etc.). Thus, secure communications are obligatory to most businesses/organizations and in this sense seen as a primordial requirement of technological and military exploits. However, as technologies continue to explode, especially with the development of modern computing technologies, the internet of things (IoT), 5G and future generation networks, adverse robust ways of information theft continue to grow \cite{jrnl_6g, physical_layer_security_iot}. In practice, a total secured communication is unattainable, nevertheless, theories seem to support a measure that is acceptable \cite{on_the_secrecy}, \cite{a_survey_on_wireless_security}. It is important that communication be unique in all the layers of the communication model - open system interconnection (OSI) and/or the internet model to guarantee its security. Different protocols and techniques have been developed in the literature for the security in the layers of these models \cite{a_survey_on_wireless_security}. Public and private key-based cryptographic security measures are most widely used in many communication systems. However, cryptographic security is heavily computation demanding in one hand, thus impractical in many IoT applications, and subject to sophisticated external attacks with the advent of modern computing facilities on the other hand. Developing novel security measures to combat such attacks is therefore of prime interest for many researchers. In this context, the notion of physical layer security has attracted significant attention due to its ability to provide information-theoretic security  \cite{a_survey_on_wireless_security, jrnl_secrecy, jrnl_secrecy_sinr, jrnl_alex_fd_sec}.

The physical layer is similar in most communication models as it deals with processing the encapsulated message for transmission via the channel \cite{physical_layer_security_5G}. In wireless communications, it deals directly with the electromagnetic waves referred to as signals. These signals can be compromised via eavesdropping and jamming of legitimate receivers. Focusing on the eavesdropping, the security in the physical layer can be subdivided into key-based and keyless security models. The primary objective of both models is to reduce the ability of an illegitimate user to gain access to the transmitted message. While the key-based models use information obscurity as its main tool, the keyless models detect the possible information leak in the presence of eavesdropper(s)\footnote{The kind of eavesdroppers referred to in this paper are considered as passive Wyner wiretappers \cite{wyner} which do not attempt to alter the transmitted message but try to overhear only.} and attempts to decrease its intercepted information. The degree of information protection in a keyless physical layer security model is measured as the secrecy capacity for delay tolerant applications and the outage probability for delay intolerant applications. To maximize the secrecy capacity, \cite{on_the_secrecy} proposed an on/off algorithm that varies the power transmitted from the source especially when the eavesdropper have better channel quality. It relied on the principle that the source knows the channel state of the eavesdropper based on inherent channel monitoring. While this scheme reduces the information content received by the legitimate receiver, it also has limited practical applications as the channel information of the eavesdropper is usually unknown. Instead of reducing the transmit power, a more sophisticated approach could be to deliberately jam the eavesdropper's channel ensuring that it receives little/no information. The major limitation of this technique is that the eavesdropper will usually operate at the same band as the legitimate receiver, hence the jamming will also affect the legitimate receiver. A combination of jamming and power variation, harnessing their gains is subsequently the bedrock of modern signal jamming techniques.

Signal jamming as a physical layer protection strategy is one of the most prominent brute-force methods of limiting the information theft in keyless physical layer security exploiting the fading characteristics of the channel \cite{physical_layer_security_5G}. It entails simultaneous transmission of a signal with similar characteristics to the genuine signal but carrying no information content to cause interference to the eavesdropper's received signal. Although this technique does not guarantee that there will be no information leakage, similar to other security techniques, it reduces the probability of successful interception thereby increasing the secrecy capacity of the end-to-end communication. While jamming poses to be an effective technique for improving secrecy, there are some critical issues that affect the effectiveness of signal jamming:
\begin{enumerate}[(a)]
    \item The degree of transmit power required to increase the secrecy capacity without adversely degrading the information content of the desired receiver,
    \item The transmitter's responses to the knowledge of the possible eavesdropper(s),
    \item The optimal location to deliver the jamming signals from.
\end{enumerate}

Researchers have since investigated these requirements independently as in \cite{physical_layer_sec}, however, the investigation of the  collective effects of (a)-(c) is of practical interest due to their inter-dependency in the context of secrecy performance. While some recent studies affirm that this technique yields improvement in the secrecy capacity, they are all based on the impractical assumption that the eavesdropper(s') location is perfectly known at the transmitter \cite{uav_traj_data}. 

With respect to the known remote eavesdropper location, mobile means of delivering the jamming signals have recently been investigated in the literature. One of the effective methods proposed is the use of an unmanned aerial vehicle (UAV) in scenarios where the nodes under consideration (the source, the main receiver and the eavesdropper) are all based on the ground. This is primarily due to its aerial radio visibility of the ground terminals, its cost efficiency and its availability for low-range applications. The applications of UAVs in communications range from their use as aerial base stations \cite{cellular_uav, energy_UAV_only_traj, uav_secured_com_JTTP}, as relay nodes \cite{discrete_t}, as access/user nodes \cite{cellular_uav, uav_traj_data} to channel estimation \cite{multi_target}, etc. Recently, with the advancement of the internet of things (IoT), network of UAVs for UAV-to-UAV communications as well as for general data transmission has also been considered \cite{network_connected_uav}.

More recently, UAVs have been deployed for assisting in secure communications between ground terminals \cite{uav_cooperative_jamming, joint_trajectory}, and to act as both relay nodes and security agents between ground terminals \cite{iterative_uav}. In \cite{securingUAV_com}, the UAV is deployed with two opposing roles namely, to establish favorable and degraded channels for the legitimate and the eavesdropping links, respectively. A separate jammer UAV has been considered in \cite{mobile_jammer} to degrade the eavesdropping channel in addition to the cooperative UAV for the legitimate channel. Subsequently, UAVs have also been used to deliver classified messages to ground terminals amidst the constraints of eavesdroppers and no-fly regions in \cite{no_fly}. Critical examination reveals that the methods used in \cite{uav_cooperative_jamming, joint_trajectory, iterative_uav, securingUAV_com, mobile_jammer, no_fly} are similar in principle since they optimized the transmitted power, the UAV jamming power and its trajectory for the corresponding scenarios. However, a strong assumption made in \cite{uav_cooperative_jamming, joint_trajectory, iterative_uav, securingUAV_com, mobile_jammer, no_fly} is that the location of the eavesdropper(s) is known to the source and/or the UAV(s). Although this assumption simplifies the respective problem in each scenario, it is grossly impractical. In most practical communication scenarios, even knowing the presence of an eavesdropper is often very difficult let alone knowing their exact locations or channel state information (CSI). This practical challenge motivates us to investigate secret communication with unknown eavesdropper location in this paper. We consider UAV-aided jamming technique for proactively degrading the eavesdropping channel at unknown ground point for improving the achievable secrecy rate.

An attempt to introduce eavesdropper obscurity has also been made by Miao Cui, et al. in \cite{robust_trajectory_power}. The authors in \cite{robust_trajectory_power} considered the UAV as the information source and optimized its trajectory and transmitting power to a legitimate receiver amidst a group of eavesdroppers located within an independent small uncertainty region. The trajectory of the UAV has been optimized to find the best points in the space to deliver the maximum information to the legitimate receiver while the eavesdroppers receive minimum information. In contrast, we consider the UAV with an opposing role in this paper to degrade the eavesdropper's channel via cooperative jamming. Note that our work differs from \cite{robust_trajectory_power} not just in terms of the UAV's role, but also in terms of guaranteed secrecy performance. In fact, the achievable secrecy performance in \cite{robust_trajectory_power} cannot be guaranteed as the uncertainty region expands and overlaps with the certainty region of the legitimate receiver. Furthermore, a network of legitimate and illegitimate UAVs has been considered in \cite{secure_uav_uav} in which a UAV acts as the base station to transmit signal to other legitimate UAVs in altitude and the eavesdropper UAVs from unknown locations try to overhear the signal. The secrecy outage probability and the average secrecy rate performance have been analyzed. Since all the nodes are at the same altitude, the gains of aerial visibility of UAV was subdued. In this work, we intend to explore this opportunity for ground nodes (source, legitimate receiver and eavesdropper) in order to maximize the benefits of aerial visibility of the UAV while constrained by the properties of ground propagation.

We formulate the problem of maximizing the average secrecy rate under the unknown eavesdropper location assumption by jointly optimizing the source transmit power, the UAV trajectory and its jamming power. The problem is strictly non-convex due to the correlation of the optimization variables in the problem. Therefore, in this work, we sequentially optimize the flight path of a UAV, its jamming power and the transmitted power by the source node to ensure secure communication in the considered scenario. One set of variables are optimized in each step while keeping the others fixed. The main contributions in this paper can be summarized as:
\begin{enumerate}[(a)]
    \item Developing the mathematical analysis of UAV-aided  jamming applications to secure wireless communication when the location of the eavesdropper is completely unknown.\label{a}
    \item Applying the block coordinate descent method and successive convex approximation (SCA) technique with the aid of the first-order Taylor series expansion.\label{b}
    \item Unveiling the influence of the unknown eavesdropper's received power on the average secrecy rate between the source and the legitimate receiver.
    \item Validating the formulations and the solutions by demonstrating the performance of the proposed algorithm against existing UAV-aided secure communication schemes through extensive numerical simulations.\label{c}
\end{enumerate}
%Based on the simulation results, we obtained better secrecy rate performance when compared with using direct UAV flight path but had reduced performance compared to known eavesdropper location. We also showed that there is an optimal flight duration of the UAV to obtain maximum average secrecy rate\footnote{Based on the set simulation parameters}. The flight altitude of the UAV plays less role in increasing average secrecy rate when compared to its speed and the SNR of ground nodes.

The rest of this paper is organized as follows: Section~\ref{sec_sysmod} describes the UAV-aided communication system model and the problem formulation. The proposed solution is developed in Section~\ref{sec_solution}. Simulation results are presented in Section~\ref{sec_sim} before making the concluding remarks in Section~\ref{sec_con}.

\begin{figure}[ht!]
\centering
\includegraphics[width=0.8\linewidth]{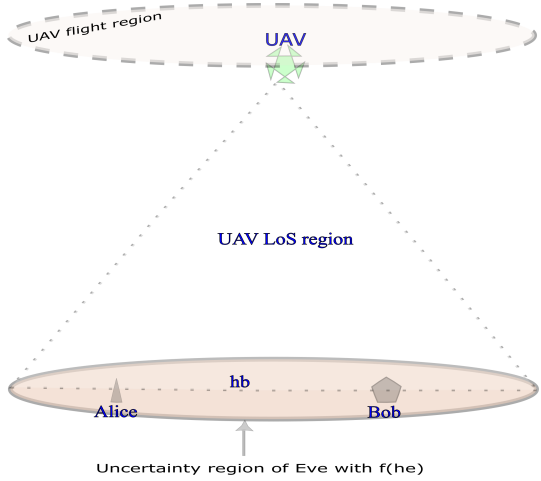}
\caption{UAV-aided jamming for secure communication.}
\label{sys_model}
\end{figure}

\section{System Model}\label{sec_sysmod}
We consider a secure wireless communication scenario between a base station (BS) acting as a transmitting source (Alice) located at an a-priori known ground point\footnote{z-coordinate represents the altitude and the ground point is located at z=0.} $\bf{w}_a=[x_a,y_a,0]^T$ and a receiver (Bob) at an a-priori known ground point $\bf{w}_b=[x_b,y_b,0]^T$ as shown in Fig.~\ref{sys_model}. However, an eavesdropper (Eve) lurks around the area in an unknown ground location, $\tilde {\bf{w}}_e=[\tilde {x}_e,\tilde {y}_e,0]^T$, but within the area where the wireless signal can be received. We denote the complex block-fading channels of Alice with Bob and Eve as $g_b$ and $g_e$, respectively. Since Eve's location is unknown, Alice's transmission power $P_a$ is a function of Bob's  channel power gain $h_b = |g_b|^2$ alone; hence $P_a=P(h_b)$, i.e. Alice varies her transmission power depending on the channel state of Bob. The average secrecy rate and secrecy capacity derived from Shannon's information content are given respectively as \cite{on_the_secrecy}
\begin{multline}\label{secrecy_rate}
R_s = \int_0^\infty\int_0^{h_b}[\underbrace{\log_2(1+h_b P(h_b)}_{\text{information rate of Bob}}  -\underbrace{\log_2(1+h_e P(h_b)}_{\text{information rate of Eve}}]^+\\ 
\times f(h_b) f(h_e) {\rm d}h_e {\rm d}h_b,
\end{multline}
\begin{flalign}\label{Cs_o}
C_s = \max_{P(h_b)} ~&~  R_s
\end{flalign}
where $R_s, ~ C_s, ~P(h_b), ~h_b ~{\textrm {and}} ~h_e = |g_e|^2$ are the average secrecy rate, secrecy capacity, transmit power from Alice, the channel power gain of Bob and Eve, respectively, and $[a]^+$ indicates $\max(0,a)$ \footnote{Nota bene: All logarithms used in this work is of base $2$ since we refer to digital communications.}. Note that $[\cdot]^+$ imposes a constraint such that Eve cannot receive higher information than Bob at any time during the communication. Accordingly, the limits of the integrals in \eqref{secrecy_rate} are defined such that when $h_e \ge h_b$, the mutual information between Alice and Eve is upper-bounded by $\log_2(1+h_b P(h_b))$. The objective of keyless physical layer security is to ensure that \eqref{Cs_o} is sustained at its optimal value over the duration of the communication.

Note that the achievable secrecy rate in \eqref{secrecy_rate} describes the secrecy rate as the difference of the average information rates of Bob and Eve over all fading realizations of Bob and Eve. The non-negativity assumption on the secrecy rate $[\cdot]^+$ requires that the location of Eve revolves around that of Bob and not beyond the coverage region of Alice. However, in practice, Eve may even be located at positions closer to Alice than Bob and thereby receive stronger signals than Bob assuming they both share the same channel model based on the proximity to the transmitter alone. In such scenarios, the achievable secrecy rate would be zero.

To ameliorate the aforementioned challenge, we deploy a UAV that will deliver jamming signals to reduce the information acquired by Eve while attempting to sustain that obtained by Bob. However, in UAV-aided communications, a common challenge is to optimize the UAV trajectory \cite{cellular_uav}. In secure communications, the challenge is further proliferated by the unknown eavesdropper location. We aim at addressing this challenge in the following sections. The UAV flight path will be optimized to ensure that for any location of Eve within the coverage region of Alice, its information rate will be continually below that of Bob, thereby, achieving positive secrecy rates. 

We assume that the UAV is not equipped with any tracking devices. Therefore, the UAV will not be able to locate or track Eve despite having a clear line-of-sight (LoS) to all points within the coverage region of Alice due to aerial visibility.

Furthermore, if the UAV flies horizontally at constant altitude from an initial point $q_0$ to a final point $q_f$, its ascent and descent flight path to the initial and final ground points can be neglected. %And greater emphasis laid on finding optimum points to deliver the jamming signals \cite{uav_cooperative_jamming}. 
The UAV flight duration, $T$, is sampled at discrete time-stamps of $N$ equal time slots with duration of $\delta=T/N$ \cite{discrete_t, uav_cooperative_jamming}. With large number of time slots $N$, we can assume that the UAV maintains constant speed $V$ m/s within a slot $\delta$ and transmits almost continuously. For simplicity, we assume that $V$ is constant over the entire flight duration as also assumed in \cite{uav_cooperative_jamming}. If the distance covered in each sample is small enough, we can assume that the UAV is stationary at each sample point. Considering a large number of sample points, the UAV is assumed to send jamming signal continuously. These sampled points can be denoted as $\boldsymbol{q}[n]=\big[x[n],y[n],z[n]\big]^T$, $n \in \{1,...,N\}$, which satisfies the following constraints:
%defining the box... store equations in a box to use later (if you do not want to refer the equation %number) instead or retyping
\newsavebox\qconst
%filling the box
\savebox\qconst{\vbox{
\begin{subequations}\label{q_c}
\begin{align}
\|\boldsymbol{q}[n+1]-\boldsymbol{q}[n]\|^2 & \leq (V\delta)^2 \label{q_c1}\\
\|\boldsymbol{q}[1]-\boldsymbol{q}_0\|^2 & \leq (V\delta)^2 \label{q_c2}\\
\boldsymbol{q}[N] & = \boldsymbol{q}_f \label{q_c3}\\
    \|\boldsymbol{q}[n]-\boldsymbol{w}_a\|+\|\boldsymbol{q}[n]-\boldsymbol{w}_b\| & \leq 2a \label{q_c4}\\
\boldsymbol{q}(x_n,y_n,z_n) &= \boldsymbol{q}(x_n,y_n,H).\label{q_c5}
\end{align}
\end{subequations}
}}
%use the box
\usebox\qconst \\
Inequalities \eqref{q_c1} and \eqref{q_c2} ensure that the distance covered by the UAV within the flight samples does not exceed the parametric distance. The velocity $V$ m/s is chosen such that the total distance covered by the UAV through the samples will be greater than or equal to the Euclidean distance between $q_0$ and $q_f$, i.e., $(V\delta) \geq \|\boldsymbol{q}_f-\boldsymbol{q}_o\|$, otherwise the system will be intractable. This ensures that the UAV travels at least in a straight path from its initial to its final points for a given total flight duration. The equality in \eqref{q_c3} ensures that the final flight point of the UAV is at the a-priori final destination, while \eqref{q_c4} allows the UAV to remain within the uncertainty region where the eavesdropper can be found. This region is postulated as an ellipse and physically represents a cellular coverage region of Alice. $a$ determines the size of the ellipse and satisfies$~\{\textrm{for}~a> \|\boldsymbol{w}_b-\boldsymbol{w}_a\|\}$, $w_a$ and $w_b$ are the two foci of the ellipse, ensuring that Bob is not a cell-edge user. Finally, \eqref{q_c5} places the UAV to fly at constant altitude denoted by $H$ meters.

Assuming the ground fading channel between Alice and Bob is Rayleigh distributed, the lower bound of the channel power gain (corresponding to the worst channel condition) with the jamming signal delivered by the UAV is given by \ref{hb} as obtained in \cite{securingUAV_com}, \cite{joint_trajectory}, \cite{mobile_jammer}. We note that \ref{hb} is the upper bound of the random complex channel $g_b$ \cite{securingUAV_com}. 
\begin{align}\label{hb}
h_b[n]=\frac {\overbrace{\beta_od_{ab}^{-\psi}}^{\text{ground channel gain}}}{\underbrace{P_u[n] \beta_od_{qb}^{-2}[n]+1}_{\text{LoS jamming signal attenuation}}},
\end{align}
where $\psi$ is the ground path loss component between Alice and Bob, $\beta_0$ represents the signal-to-noise ratio (SNR) at a reference distance $(d_0=1)$m of the ground channels, $d_{ab}$ and $d_{qb}$ are the Euclidean distance between Alice, UAV and Bob respectively and $P_u$ is the UAV jamming signal power. To ensure that the power levels of the communication is within acceptable range, $P_u$  and $P_a$ are subjected to average and peak power constraints described as:
\newsavebox\pconst
%filling the box
\savebox\pconst{\vbox{
\begin{subequations} \label{P_c}
\begin{align}
0 \leq P_u[n] \leq P_{umax} \label{P_cpeak_pu}\\
\frac{1}{N} \sum_{n=1}^N P_u[n] \leq \bar{P}_{ub} \label{P_cav_pu}\\
0 \leq P_a[n] \leq P_{amax} \label{P_cpeak_pa}\\
\frac{1}{N} \sum_{n=1}^N P_a[n] \leq \bar{P}_{ab}. \label{P_cav_pa}
\end{align}
\end{subequations}
}}
\usebox\pconst

\subsection{Problem Formulation}\label{sec_pro}
Let $\boldsymbol {\rm Q}=\{q[n], n\in N\}$, $\boldsymbol {\rm p}_a=\{P_a[n], n\in N\}$, and $\boldsymbol {\rm p}_u=\{P_u[n], n\in N\}$ be the set of UAV sample points (representing its trajectory when connected by a straight line), the set of power transmitted by Alice as well as the UAV, respectively. We aim at solving \eqref{Cs_o} by alternatingly optimizing $\boldsymbol {Q}$, $\boldsymbol {p}_a$ and $\boldsymbol {p}_u$. Using Reimann sum and averaging through all fading realizations of $h_b$, we approximate \eqref{secrecy_rate} to obtain
\begin{multline}\label{Rsecrecy_rate}
R_s = \frac{1}{N}\sum_{n=1}^N \int_0^{h_b[n]}[\log(1+h_b[n]P_a[n]\\
-\log(1+h_e[n]P_a[n]] f(h_e) dh_e.
\end{multline}
%\begin{equation*}
%\textrm{s.t.} \quad \eqref{q_c} \quad \textrm{and} \quad \eqref{P_c}
%\end{equation*}
To solve \eqref{Rsecrecy_rate}, we need to know the possible distribution of the fading channel of Eve which can be obtained by historical measurements collected over the region covered by Alice (represented in this model as an ellipsis as in \eqref{q_c4}). Using the Rayleigh fading realization assumption, we have its squared envelop as
\begin{equation}
f(h_e)=\frac{1}{y_e} e^{-\frac{h_e[n]}{y_e}}, \label{fhe}
\end{equation}
where $y_e$ is the average received envelop power of Eve and can be obtained via measurements and/or estimations. Substituting \eqref{fhe} in \eqref{Rsecrecy_rate} and solving the integral, we obtain
\begin{multline}\label{Rs}
R_s=\frac{1}{N}\sum_{n=1}^N \underbrace{\log(1+h_b[n]P_a[n])}_{\text{information rate of Bob}}\\
-\underbrace{\int_0^{h_b[n]}\frac {P_a[n]e^{-\frac{h_e[n]}{y_e}}}{1+h_e[n]P_a[n]} dh_e}_{\text{information rate of Eve}}.
\end{multline}
%\begin{equation*}
%\textrm{s.t.} \quad \eqref{q_c} \quad \textrm{and} \quad \eqref{P_c}
%\end{equation*}
The secrecy rate in \eqref{Rs} can be further simplified as \cite[eq. 3.352.1]{tisp}
\begin{multline}\label{mRs}
R_s=\frac{1}{N}\sum_{n=1}^N \log(1+h_b[n]P_a[n])\\
-e^{\frac{1}{y_eP_a[n]}}\left [E_i\left(-\frac{h_b[n]}{y_e}-\frac{1}{y_eP_a[n]}\right) \right.\\
\left. -E_i\left(-\frac{1}{y_eP_a[n]}\right)\right],
\end{multline}
%\begin{equation*}
%\textrm{s.t.} \quad \eqref{q_c} \quad \textrm{and} \quad \eqref{P_c}
%\end{equation*}
where $$E_i(x)=\int_x^\infty \frac{e^{-t}}{t}dt$$ is the exponential integral. 
We note that \eqref{Rs} is equivalent to \eqref{mRs} and they can be used interchangeably depending on the parameter been inferred. Thus we substitute the objective function in \eqref{Cs_o} with the elaborated form in \eqref{mRs} to obtain the following optimization problem\footnote{We neglected the constant scaling factor $\frac{1}{N}$ in the objective function as this does not affect the optimal solution.}
\begin{subequations}\label{P1_mRs}
\begin{align}
(P1): \max_{\boldsymbol {\rm p}_a,\boldsymbol {\rm p}_u,\boldsymbol {\rm Q}} ~&~ \sum_{n=1}^N \log(1+h_b[n]P_a[n]) \nonumber\\
& -e^{\frac{1}{y_eP_a[n]}}\left [E_i\left(-\frac{h_b[n]}{y_e}-\frac{1}{y_eP_a[n]}\right) \right. \nonumber\\
&\left. -E_i\left(-\frac{1}{y_eP_a[n]}\right)\right] \label{P1_o}\\
\textrm{s.t.} ~&~ \|\boldsymbol{q}[n+1]-\boldsymbol{q}[n]\|^2 \leq (V\delta)^2 \label{P1_c1}\\
& \|\boldsymbol{q}[1]-\boldsymbol{q}_0\|^2 \leq (V\delta)^2 \label{P1_c2}\\
& \boldsymbol{q}[N]  = \boldsymbol{q}_f \label{P1_c3}\\
& \|\boldsymbol{q}[n]-\boldsymbol{w}_a\|+\|\boldsymbol{q}[n]-\boldsymbol{w}_b\|  \leq 2a \label{P1_c4}\\
& \boldsymbol{q}(x_n,y_n,z_n) = \boldsymbol{q}(x_n,y_n,H),\label{P1_c5}\\
& 0 \leq P_u[n] \leq P_{umax} \label{P1_c6}\\
& \frac{1}{N} \sum_{n=1}^N P_u[n] \leq \bar{P}_{ub} \label{P1_c7}\\
& 0 \leq P_a[n] \leq P_{amax} \label{P1_c8}\\
& \frac{1}{N} \sum_{n=1}^N P_a[n] \leq \bar{P}_{ab}. \label{P1_c9}
\end{align}
\end{subequations}
Problem (P1) entails that the secrecy capacity of the proposed system depends on the optimal transmission power of Alice, the jamming power delivered by the UAV and the UAV location. Unfortunately, (P1) is a non-convex optimization problem with respect to the optimization variables $(\boldsymbol {\rm p}_a,\boldsymbol {\rm p}_u,\boldsymbol {\rm Q})$ and cannot be easily solved directly. However, using a sequential and iterative technique under a block coordinate approach, we can obtain suboptimal solutions that satisfy the constraints in $\quad \eqref{q_c} \quad \textrm{and} \quad \eqref{P_c}$.

\section{Proposed Solution}\label{sec_solution}
We propose solving the non-convex problem (P1) in an alternating fashion. The proposed solution involves decomposing the original problem (P1) into three sub-problems each  characterizing a set of optimization variables. In each sub-problem, we optimize one set of variables while fixing the other variables in each iteration. The results obtained from each iteration step are analyzed with the objective value of (P1) and the iteration stops at the point when the objective value (P1) converges. %The sub-optimal results do not follow any know incremental algorithm (e.g. Newton step etc.) due to the inter-dependence of the secrecy rate on all the optimization variables, hence, the chosen exit criteria.

\subsection{Optimizing the Source Power ($P_a$)}\label{sec_Pa}
We first optimize Alice's transmit power for arbitrary initial trajectory and jamming power. Replacing the objective in problem (P1) with \eqref{Rs}, problem (P1) can be reformulated for any given $\boldsymbol {\rm Q}$ and $\boldsymbol {\rm p}_u$ as problem (P2):
\begin{subequations}\label{P2_mRs}
\begin{align}
(P2): \max_{\boldsymbol {\rm p}_a} ~&~ \sum_{n=1}^N \log(1+h_b[n]P_a[n]) \nonumber\\
~&~ \qquad-\int_0^{h_b[n]}\frac {P_a[n]e^{-\frac{h_e[n]}{y_e}}}{1+h_e[n]P_a[n]} dh_e\\
\textrm{s.t.} ~&~ \eqref{P_cpeak_pa} \quad \textrm{and} \quad \eqref{P_cav_pa}. 
\end{align}
\end{subequations}
Note that problem (P2) is still non-convex over the entire domain of $\boldsymbol {\rm p}_a$. However, for the region under peak and average power constraints, the objective can be shown to be the sum of a concave and a convex functions. The proof is relegated to Appendix~\ref{Appendix_A}.

Since the objective function of problem (P2) is differentiable (as demonstrated in Appendix~\ref{Appendix_A}), it can be solved using the Karush-Kuhn-Tucker (KKT) conditions for non-convex problems \cite{boyd_convex}:
\begin{subequations}\label{KKT}
\begin{align}
\nabla f_0(x^*)+\lambda^*\nabla f_n(x^*) & = 0,\\
\lambda^*f_n(x^*) &= 0,
\end{align}
\end{subequations}
where $f_0$ is the objective in problem (P2), $f_n$ are the constraints in \eqref{P_cpeak_pa} and \eqref{P_cav_pa} and $x^*$ is the optimal value of $P_a$. Simultaneously solving \eqref{KKT} using \cite[eq. 0.410 and 3.462.17]{tisp} respectively, we obtain
\begin{multline}\label{P2_sol}
%\frac{-\nabla f_o(x^*)}{\nabla f_n(x^*)}f_n(x^*)=0\\
-\left[\frac{h_b[n]}{1+h_b[n]P_a[n]}-\frac{1}{y_e(P_a[n])^2}e^{\frac{1}{y_eP_a[n]}}\right.\\
\left.\left[\Gamma\left(-1,{\frac{1}{y_eP_a[n]}}\right)-\Gamma\left(-1,{\frac{h_b[n]}{y_e}+\frac{1}{y_eP_a[n]}}\right)\right]\right]\\
\times \sum_{n=1}^NP_a[n]-\frac{1}{N}\bar{P}_{ab}=0,
\end{multline}
where $\Gamma(-i,z)=\frac{(-1)^i}{i!}(E_1(z)-e^{-z}\sum_{k=0}^{i-1}\frac{(-1)^kk!}{z^{k+1}})$ \cite[eq. 8.4.15]{nist}). Solving \eqref{P2_sol} with a non-linear solver produces the suboptimal values of $P_a$.

\subsection{Optimizing the UAV Jamming Power ($P_u$)}\label{sec_Pu}
To optimize the jamming power $\boldsymbol {\rm p}_u$ delivered by the UAV, we consider $\boldsymbol {\rm p}_u$ as the optimization variable while fixing the values of $\boldsymbol {\rm p}_a \quad {\rm and} \quad \boldsymbol {\rm Q}$. Problem (P1) is then reformulated while substituting for $h_b[n]$ as
\begin{subequations}\label{P3_mRs}
\begin{align}
(P3): \max_{\boldsymbol {\rm p}_u} ~&~ \sum_{n=1}^N \log \left(1+\frac{\beta_o d_{ab}^{-\psi}P_a[n]}{P_u[n]\beta_od_{qb}^{-2}[n]+1}\right)\nonumber\\
-e^{\frac{1}{y_eP_a[n]}}\times & \left[E_i\left(-\frac{\frac {\beta_od_{ab}^{-\psi}}{P_u[n] \beta_od_{qb}^{-2}[n]+1}}{y_e}-\frac{1}{y_eP_a[n]}\right)\right.\nonumber\\
&\left.-E_i\left(-\frac{1}{y_eP_a[n]}\right)\right]\\
\textrm{s.t.} ~&~ \eqref{P_cpeak_pu} \quad \textrm{and} \quad \eqref{P_cav_pu}.
\end{align}
\end{subequations}
Under the constraints, the objective of Problem (P3) is a non-convex function with respect to $\boldsymbol {\rm p}_u$ due to the non-convexity of the information rate of the Eve. However, the information rate of Bob is concave with respect to $\boldsymbol {\rm p}_u$. Hence, problem (P3) is solved using successive convex approximation (SCA). Given an initial UAV jamming power in the $k$-th iteration as $\boldsymbol {\rm p}_u^k =\{P_u^k[n],n\in N\}$; we have using first order Taylor expansion that
\begin{multline}\label{Pu_taylor}
e^{\frac{1}{y_eP_a[n]}}\left[E_i\left(-\frac{\frac {\beta_od_{ab}^{-\psi}}{P_u[n] \beta_od_{qb}^{-2}[n]+1}}{y_e}-\frac{1}{y_eP_a[n]}\right)\right.\\
\left. -E_i\left(-\frac{1}{y_eP_a[n]}\right)\right]\leq G_k[n]+T_k[n](P_u[n]-P_u^k[n]),
\end{multline}
where \begin{multline*}
G_k[n]=e^{\frac{1}{y_eP_a[n]}}\left[E_i\left(-\frac{\frac{\beta_od_{ab}^{-\psi}}{P_u^k[n]\beta_od_{qb}^{-2}[n]+1}}{y_e}-\frac{1}{y_eP_a[n]}\right)\right.\\
\left. -E_i\left(-\frac{1}{y_eP_a[n]}\right)\right]
\end{multline*}
and $T_k[n]=\frac{P_a[n]\beta_o^2d_{ab}^{-\psi} d_{qb}^{-2}[n]e^{-({\frac{\beta_od_{ab}^{-\psi}}{y_e\beta_od_{qb}^{-2}[n]P_u^k[n]+y_e}})}}{(\beta_od_{qb}^{-2}[n]P_u^k[n]+1)(P_a[n]\beta_od_{ab}^{-\psi}+\beta_od_{qb}^{-2}[n]P_u^k[n]+1)}$. 
Taking only the non-constant terms in \eqref{Pu_taylor}, problem (P3) can be reformulated as
\begin{subequations}\label{P3b_mRs}
\begin{align}
(P3b): \max_{\boldsymbol {\rm p}_u} ~&~ \sum_{n=1}^N\left[\log\left(1+\frac{\beta_od_{ab}^{-\psi}P_a[n]}{P_u[n]\beta_od_{qb}^{-2}[n]+1}\right)\right.\nonumber\\
&\qquad\qquad \left. -T_k[n]P_u[n]\right]\\
\textrm{s.t.} ~&~ \eqref{P_cpeak_pu} \quad \textrm{and} \quad \eqref{P_cav_pu}.
\end{align}
\end{subequations}
Problem (P3b) is a convex problem within the constrained region and can be efficiently solved using interior-point method or a convex solver such as CVX \cite{cvx1, cvx2}.

\subsection{Optimizing the UAV Trajectory ($\bf Q$)} \label{sec_Q}
In this sub-problem, the problem (P1) is recast to ensure that only the UAV trajectory, $\boldsymbol {\rm Q}$ is the optimization parameter. However, the reformulated problem is non-convex in $\bf Q$. Hence, to reduce computational complexity, we introduce a slack variable $\boldsymbol{M}=\{m[n]=\|q[n]-w_b\|^2, n\in N\}$ such that $d_{qb}^{-2}[n]=\frac{1}{m[n]}$. Thus we obtain the following optimization problem:
\begin{subequations}\label{P4_mRs}
\begin{align}
(P4): \max_{\boldsymbol{\rm Q},\boldsymbol{\rm M}}m~&~ \sum_{n=1}^N \log\left(1+\frac{\beta_od_{ab}^{-\psi}P_a[n]}{\frac{P_u[n]\beta_o}{m[n]}+1}\right)-e^{\frac{1}{y_eP_a[n]}}\nonumber\\
& \times \left[E_i\left(-\frac{\frac {\beta_od_{ab}^{-\psi}}{\frac{P_u[n] \beta_o}{m[n]}+1}}{y_e}-\frac{1}{y_eP_a[n]}\right)\right.\nonumber\\
& \qquad\qquad \left.-E_i\left(-\frac{1}{y_eP_a[n]}\right)\right]\\
 \textrm{s.t.} ~&~ m[n]-\|q[n]-w_b\|^2 \leq 0, \label{slack1}\\
& \text{and} \quad \eqref{q_c}.
\end{align}
\end{subequations}
Due to the non-convexity of problem (P4) with respect to the trajectory, $q[n]$, we reformulate the problem using successive approximation with the first order Taylor expansion. Let $Q_k[n]=\{q^k[n], n\in N\}$ denote the initial UAV trajectory for the $k$th iteration. Then the objective function of problem (P4) can be rewritten as 
%\begin{subequations}
\begin{multline}\label{Q_taylor}
%\begin{split}
e^{\frac{1}{y_eP_a[n]}}\left[E_i\left(-\frac{\frac {\beta_od_{ab}^{-\psi}}{\frac{P_u[n] \beta_o}{m[n]}+1}}{y_e}-\frac{1}{y_eP_a[n]})-E_i(-\frac{1}{y_eP_a[n]}\right)\right]\\ %\notag\\
\leq O_k[n]+W_k[n](q[n]-q^k[n])
\end{multline}
\begin{align}
-\|q[n]-w_b\|^2 \leq S^k[n],
\end{align}
%\end{subequations}
where \begin{multline*}
O_k[n]=e^{\frac{1}{y_eP_a[n]}}\left[E_i\left(-\frac{\frac {\beta_od_{ab}^{-\psi}}{\frac{P_u[n] \beta_o}{m_k[n]}+1}}{y_e}-\frac{1}{y_eP_a[n]}\right)\right.\\
\left. -E_i\left(-\frac{1}{y_eP_a[n]}\right)\right]
\end{multline*}
%\begin{multline*}
$$W_k[n]=\frac{\beta_o^2d_{ab}^{-\psi}P_u[n]e^-{\frac{\beta_od_{ab}^{-\psi}}{y_e\left(1+\frac{\beta_oP_u[n]}{m_k[n]}\right)}}}
{y_e\left(-\frac{1}{y_eP_a[n]}-\frac{\beta_od_{ab}^{-\psi}}{y_e\left(1+\frac{\beta_oP_u[n]}{m_k[n]}\right)}\right)\left(1+\frac{\beta_oP_u[n]}{m_k[n]}\right)m_k^2[n]},
$$%\end{multline*}
and $S^k[n]=\|q_k[n]\|^2-2[q_k[n]-w_b]^Tq[n]-\|w_b\|^2$.
Under similar conditions as of problem (P3), (P4) can be reformulated as
\begin{subequations}\label{P4b_mRs}
\begin{align}
(P4b): \max_{\boldsymbol{\rm Q},\boldsymbol{\rm M}} ~&~ \sum_{n=1}^N \log\left(1+\frac{\beta_od_{ab}^{-\psi}P_a[n]}{\frac{P_u[n]\beta_o}{m[n]}+1}\right)-W_k[n]m[n]\\
\textrm{s.t.} ~&~ m[n]+S^k[n] \leq 0,\\
~&~ \text{and} \quad \eqref{q_c}.
\end{align}
\end{subequations}
Problem (P4b) is a convex problem in $\boldsymbol {\rm Q}$ under the specified constraints and can be solved using interior-point methods or with a convex solver. The overall procedure has been summarized in Algorithm~1.
\begin{algorithm} [!ht]\label{algo1}
    \caption{Iterative algorithm for solving $\boldsymbol {\rm p}_a,~\boldsymbol {\rm p}_u,~\textrm{and}~\boldsymbol {\rm Q}$}
  \begin{algorithmic}[1]
    \STATE Initialize $\boldsymbol {\rm p}_u~\textrm{and}~\boldsymbol {\rm Q}$ such that the constraints in \eqref{P_cpeak_pu}, \eqref{P_cav_pu} and \eqref{q_c} are satisfied.
    \STATE $m \leftarrow 1.$
    \STATE \textbf{repeat}
    \STATE Compute and update $\boldsymbol {\rm p}_a$ in \eqref{P2_sol} with given $\boldsymbol {\rm p}_u$ and $\boldsymbol {\rm Q}$.
    \STATE Using updated $\boldsymbol {\rm p}_a$ and current $\boldsymbol {\rm Q}$, solve \eqref{P3b_mRs} for $\boldsymbol {\rm p}_u$.
    \STATE With given $\boldsymbol {\rm p}_a$ and $\boldsymbol {\rm p}_u$, find $\boldsymbol {\rm Q}$ by solving problem \eqref{P4b_mRs}.
    \STATE Compute $R_s$ as defined in \eqref{mRs}.
    \STATE $e=\frac{R_s^{new}-R_s^{old}}{R_s^{new}}$.
    \STATE $m \leftarrow m + 1.$
    \STATE \textbf{until} {$e< \theta$ OR $m\geq m_{max}$.}
    \STATE \textbf{Output:} $\boldsymbol {\rm p}_a,~\boldsymbol {\rm p}_u,~\textrm{and}~\boldsymbol {\rm Q}$.
  \end{algorithmic}
\end{algorithm}

\section{Simulation Results and analysis}\label{sec_sim}
In this section, we evaluate the performance of the proposed solution approach through numerical simulations. We implement the solution discussed in Section \ref{sec_solution} following the procedure described in Algorithm 1. The optimization parameters are initialized by solving the feasibility problem such that the the initial values just satisfy their respective constraints. The feasibility problem can be formulated by setting the objective of problem \eqref{P1_mRs} to zero, with all the primary constraints unchanged. %, I solved with cvx to get the values of Pa, PS and Q that satisfied the constraints. 
Then, by iteratively optimizing each parameter with the knowledge of the others, we obtain the suboptimal solution to problem (P1) when the error ($e$) between steps is less than $\theta$ (where $\theta=10^{-5}$) or the maximum number of iterations is reached (where $m_{max}=200$). 

\begin{figure}[!tbp]
\centering
\includegraphics[width=\linewidth]{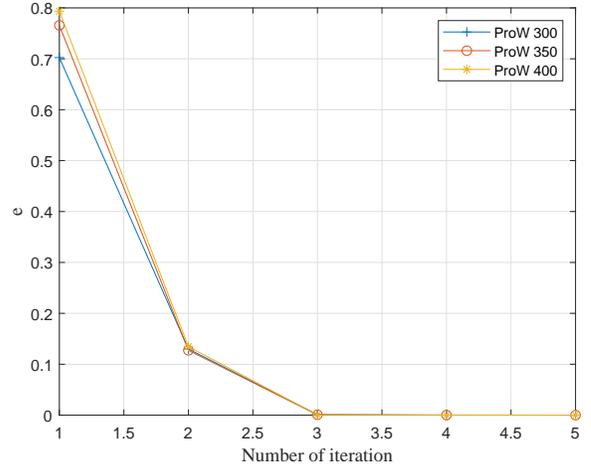}
\caption{Convergence performance of Algorithm~1.}
\label{conv}
\end{figure}

Similar to the convergence analysis in \cite{joint_trajectory}, Algorithm~1 is guaranteed to converge for all feasible initial points. This is shown in Fig.~\ref{conv} where a fast convergence is observed for different scenarios of the UAV flight time. In Fig.~\ref{conv}, the ProW algorithm represents the proposed solution to the unknown Eve problem while the associated numbers represent the UAV flight time. In all the simulations, we used the parameters as described in Table~\ref{table_par} unless otherwise specified.
\begin{table}[!ht]
\begin{center}
\caption{Simulation Parameters}\label{table_par}
    \begin{tabular}{ | p{3cm} | l | p{3cm} |}
    \hline
    %\multicolumn{3}{|c|}{\textbf{Table I:} Simulation Parameters} \\
    %\hline
    \textbf{Simulation parameter} & \textbf{Symbol} & \textbf{Value} \\ \hline
    Alice location & $\boldsymbol{w}_a$ & $[0,0,0]$ \\ \hline
    Bob location & $\boldsymbol{w}_b$ & $[300,0,0]$ \\ \hline
    Eve location & $\boldsymbol{w}_e$ & $[150,250,0]$, $[350,0,0]$, $[300,20,0]$, $[300,70,0]$ \\ \hline
    Initial UAV location & $\boldsymbol{q}_o$ & $[-100,100,H]$ \\ \hline
    Final UAV location & $\boldsymbol{q}_f$ & $[500, 100,H]$ \\ \hline
    UAV height(when fixed) & $H$ & $100$m \cite{uav_cooperative_jamming}\\ \hline
    Velocity per sample(when fixed) & $V$ & $3$m/s \cite{uav_cooperative_jamming}\\ \hline
    Duration per sample(when fixed) & $\delta$ & $0.5$s \cite{uav_cooperative_jamming}\\ \hline
    Signal-to-noise ratio & $\beta_o$ & $90$dB \cite{uav_cooperative_jamming}\\ 
    \hline  %1e9
    Average received envelop power & $y_e$ & $20$dBm\\ \hline%2W
    Average UAV transmit power & $\bar{P}_{ub}$ & $10$dBm \cite{uav_cooperative_jamming}\\ \hline%0.1W
    Maximum UAV power & $P_{umax}$ & $4$Pub \cite{uav_cooperative_jamming}\\ \hline%0.039810717055W
    Average Source power & $\bar{P}_{ab}$ & $30$dBm \cite{uav_cooperative_jamming}\\ \hline%1W
    Maximum source power & $P_{amax}$ & $36$dBm \cite{uav_cooperative_jamming}\\ \hline%3.9810717055W
    Radius of uncertainty region (when fixed) & $a$ & $450$m\\ \hline
    Path loss for ground communication (urban area cellular radio) & $u$ & $3.4$\\ \hline
    \end{tabular}
\end{center}
\end{table}

\begin{figure}[!tbp]
\centering
\includegraphics[width=\linewidth]{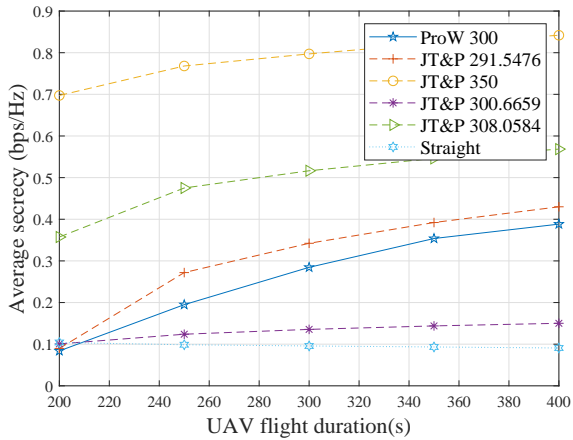}
\caption{Average secrecy rate with `unknown' as well as `known' eavesdropper locations, and direct UAV flight path.}
\label{secrecy}
\end{figure}

\begin{figure}[!tbp]
\centering
\includegraphics[width=\linewidth]{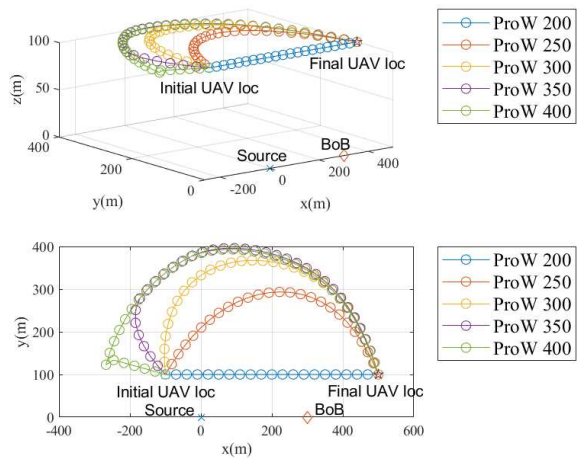}
\caption{UAV flight trajectory in 2D and 3D view while Eve location is unknown (For clarity, we use $\delta=10$).}
\label{UAV_trajectory}
\end{figure}

\begin{figure}[!tbp]
\centering
\includegraphics[width=\linewidth]{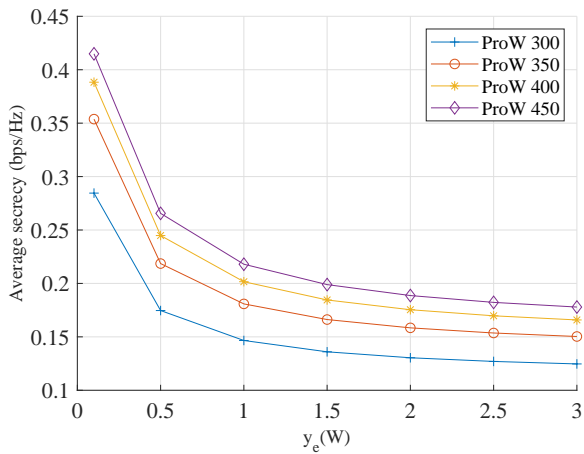}
\caption{Effect of average received envelop power of Eve on average secrecy rate.}
\label{eve_envelope}
\end{figure}

\begin{figure}[!tbp]
\centering
\includegraphics[width=\linewidth]{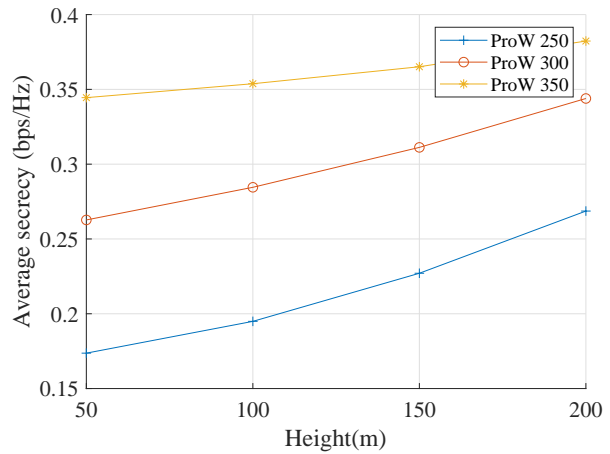}
\caption{Influence of UAV altitude (height) on average secrecy rate under the proposed scheme.}
\label{height}
\end{figure}

\begin{figure}[!tbp]
\centering
\includegraphics[width=\linewidth]{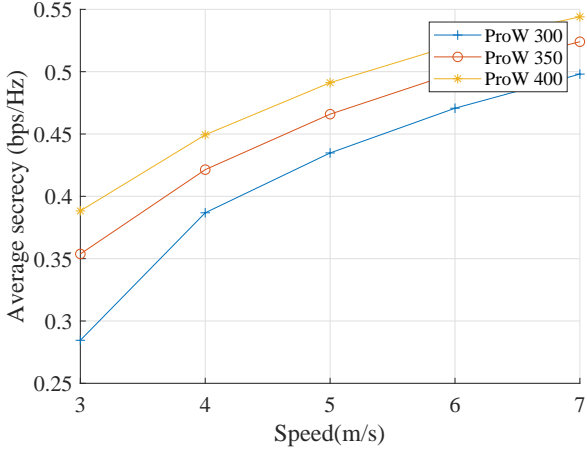}
\caption{Influence of UAV flying speed on average secrecy rate with obscure Eve.}
\label{speed}
\end{figure}

\begin{figure}[!tbp]
\centering
\includegraphics[width=\linewidth]{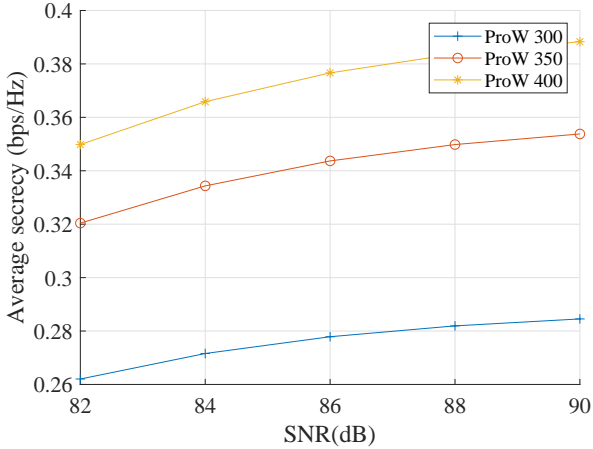}
\caption{Average secrecy rate versus signal-noise-ratio (SNR) with obscure Eve.}
\label{snr}
\end{figure}

We then analyze the secrecy rate performance of the proposed scheme as compared with the existing schemes. In Fig.~\ref{secrecy}, the performance of the unknown Eve location scenario using the proposed joint trajectory and power optimization algorithm (referred to as ProW) is compared the known Eve location scenario considered in \cite{uav_cooperative_jamming} (referred to as JT\&P). We also compare the performance with the baseline scheme without optimizing the UAV trajectory referred to as Straight), in which the UAV flies straight to the location above the eavesdropper. The associated numbers in the legends represent the respective Euclidean distances from Alice (source) to Bob for the ProW algorithm (recall unknown Eve location) and from Alice to Eve for the JT\&P algorithm based on the locations specified in Table~\ref{table_par}. Nevertheless, from Fig.~\ref{UAV_trajectory} to Fig.~\ref{snr}, the numbers attached to the acronyms depict the UAV flight time in seconds. Results in Fig.~\ref{secrecy} illustrate that the direct flight path with constant power (Straight) scheme performs the worst in terms of the average secrecy rate. %, especially, as the sample points of the UAV increases with UAV flight duration. 
Due to the jamming signals delivered by the UAV, the average secrecy rate of the JT\&P scheme is zero when the Eve is at the same location as Bob. As Eve moves away from Bob, the average secrecy rate increases since the UAV locates Eve and stays at an optimum location to jam her signal. Nevertheless, since the location of Eve is unknown (as in ProW 300), the average secrecy rate is shown to be close to the JT\&P scheme when Eve is closer to Alice and is supposed to receive more information content without the UAV jamming.
%Nevertheless, the unknown Eve location scenario performance is below that of when Eve's location is known and located far from the main receiver (Bob). This is primarily due to the inability of the UAV to locate Eve, ensuring that its flight points may sometimes be far from Eve; thereby decreasing the average secrecy rate.
However, considering that ProW 300 is near to a  practical scenario, this marginal decrease in performance may be considered as the near practical trade-off to the scheme.

Also, it is important to note that the information rate of both Bob and Eve is affected by the jamming signal of the UAV. However, Eve is affected more, even when it has better channel condition (measured in terms of its average received envelope power), as the UAV regularly finds paths such that it stays further from Bob and estimates as close to Eve as possible until it flies to its final point. This allows for positive average secrecy rates shown in Fig.~\ref{secrecy}.
%1
%2
%3

The flight trajectory of the UAV with respect to Alice and Bob is shown in Fig.~\ref{UAV_trajectory}. The 2D plot shows that from an aerial view, the trajectory of the UAV follows a given pattern bound by the uncertainty region of Eve (ellipse) provided it flies at a constant altitude. However, to demonstrate the effectiveness of the proposed approach in practice, we also plot a 3D view of the trajectory in Fig.~\ref{UAV_trajectory}. The 3D plot shows that the UAV trajectory moves towards the opposite of Bob while ensuring that the jamming signal is still delivered to all points within the constrained region of Eve.

%It is also interesting to note that ProW shows better performance to T\&nP even when Eve has better channel but located away from Bob due to the influence of the jamming signal delivered by the UAV. As observed in \cite{uav_cooperative_jamming}, the average secrecy rate increases exponentially with increase in the UAV flight time.
%4
%5
Furthermore, Fig.~\ref{eve_envelope} examines the constraints posed by the assumptions on the property of Eve's channel. 
We recall that the only known property of Eve is its average received envelop power, $y_e$. Hence Fig.~\ref{eve_envelope} presents the effect of varying $y_e$ on the average secrecy rate. It can be observed that increase in $y_e$ decreases the average secrecy rate via a positive exponential path. Hence, for large values of $y_e$ characterizing Eve having better reception equipment and channel state as compared to Bob, the decrement in average secrecy rate with respect to increasing $y_e$ becomes negligible. The optimized UAV path ensures that even when the location of Eve is unknown, the average secrecy rate of the communication between Alice and Bob can be guaranteed despite Eve supposedly receiving signals with high envelope power. While this average secrecy rate is low, it can be improved by increasing the time of flight of the UAV or allowing the UAV to fly throughout the communication duration as shown in Figs. \ref{secrecy} and \ref{eve_envelope}. %and discussed in \cite{uav_cooperative_jamming}, \cite{uav_secured_com_JTTP}, \cite{uav_traj_data}. 

Other factors that affect the average secrecy rate in  the considered scenario of unknown eavesdropper location include the UAV height and speed, and the SNR of the environment.
%The information rate of Eve increases with its proximity to the source which acts as a measure of its better channel quality. However, the rate is bound by the information rate of Bob satisfying the assumption of variable rate scheme.
Figs.~\ref{height},~\ref{speed} and~\ref{snr} show the average secrecy rate of the proposed system compared to the UAV altitude, speed and ground node SNR, respectively. It can be observed that the impact of flying the UAV at higher altitude is minimal in terms of average secrecy rate, compared to allowing it to fly at longer duration. The trend in Fig. \ref{height} suggests that the average secrecy rate increases with increase in the UAV altitude/height. However, we observed from our simulations that for large values of UAV flight altitude, the trajectory optimization problem (P4) becomes infeasible. Increasing the UAV speed and ground nodes SNR tend to increase the average secrecy rate with a logarithmic path. Nevertheless, the rate of increase is higher with the UAV speed than the SNR. As the UAV speed increases, its sample points increase allowing it to deliver more jamming signal to Eve within its flight time. Similar to results observed in \cite{secure_uav_uav}, increasing the ground SNR improves the secrecy, however, this parameter is subject to characteristics of the outdoor environment which cannot be easily controlled.
%figures are located in C:/Users/obinn/OneDrive/Desktop/UAV_Latex/Work1/graphs_fig/
%6
%7
%8
\begin{figure}[!tbp]
\centering
\includegraphics[width=\linewidth]{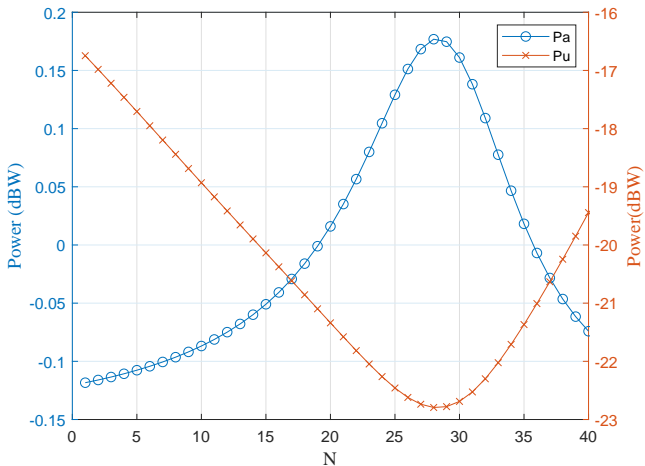}
\caption{Comparing transmitted power between Alice and UAV.}
\label{power}
\end{figure}

The optimized transmitting power of Alice $(P_a)$ and the optimized UAV jamming power $(P_u)$ are plotted in Fig. \ref{power}. While Alice transmits at its maximum power when the UAV is close to Bob, the UAV transmits minimum jamming signal. This ensures that the UAV interference to Bob is minimal and Bob continues to receive the information sent by Alice.

\section{Conclusion}\label{sec_con}
In this paper, we have exploited UAV-aided jamming technique in reducing the information rate received by an eavesdropper in an unknown location. We solved the achievable secrecy rate maximization problem using sequential block coordinate optimization method. While we were constrained by the elusive nature of the eavesdropper location, we obtained a secrecy rate that is comparable to a scenario when the eavesdropper's location is known. We also showed that the UAV speed and flight duration are amongst the main parameters to consider while using UAV to increase physical layer security. Most importantly, we have demonstrated that the average received envelope power of the eavesdropper cannot guarantee better information content as the secrecy rate tends to stabilize with large envelope power. We propose that future works investigate predicting the eavesdropper location with the aid of deep learning techniques in order to update the UAV flight path in real-time. This could reduce the latency in continuously solving the optimization problem for each communication block.

\appendices
\section{}\label{Appendix_A}
In this section, we show that the non-convexity of \eqref{Rs} is the sum of a concave and a convex functions in terms of $P_a$. From \eqref{Rs}, we obtain
\begin{multline}\label{A_Rs}
R_s=\sum_{n=1}^N \underbrace{\log(1+h_b[n]P_a[n])}_{f_1(P_a)}\\
-\underbrace{\int_0^{h_b[n]}\frac {P_a[n]e^{-\frac{h_e[n]}{y_e}}}{1+h_e[n]P_a[n]} dh_e}_{f_2(P_a)}.
\end{multline}
We consider \eqref{A_Rs} in two parts, showing their convexity with the second derivative method. In general, the convexity of a function is defined as \cite{boyd_convex}
\[f^{''}(x) = \left\{
  \begin{array}{lr}
    Convex & : > 0\\
    Concave & : < 0\\
    Affine &: = 0.
  \end{array}
\right.
\]
Thus we have from \eqref{A_Rs} that  $$f_1^{''}(P_a)=-\left(\frac{h_b}{1+h_bP_a}\right)^2.$$

We then show the convexity of the $f_2(P_a)$ using the principle that the nonnegative weighted-sum of a convex (concave) function is a convex (concave) \cite[Section 3.2.1]{boyd_convex}. The second part can be rewritten as
\begin{multline*}
\int_0^{h_b[n]}e^{-\frac{h_e[n]}{y_e}}\frac {P_a[n]}{1+h_e[n]P_a[n]}dh_e \\
\equiv \int_0^{h_b[n]}w(h_e)f(P_a,h_e)dh_e.
\end{multline*}
It has been shown in \cite{boyd_convex} that if $f(P_a,h_e)$ is convex (concave), then $f_2(P_a)$ is convex (concave). Thus, we have that the second derivative of $f_2(P_a)$ as $$f_2^{''}(P_a)=-\frac{2h_e}{(1+h_eP_a)}.$$
Thus both parts of \eqref{A_Rs} are concave functions independently under the constraint of $h_b\geq 0$ and $P_a\geq 0$. These are the positive semi-definite constraints that guarantees communication between the source and the destination. If $h_b < 0$ and/or $P_a < 0$ then no information could be transmitted successfully. Therefore, we have that \eqref{A_Rs} is the sum of a concave and a convex function ($-f(x)=\textrm{convex if} ~f(x)=\textrm{concave}$) in terms of $P_a$.

\bibliographystyle{IEEEtran}
\footnotesize{

\bibliography{IEEEabrv,refUAV}}%
\end{document}